\documentclass[aps,pra,twocolumn,showpacs]{revtex4-1}
\pdfoutput=1
\usepackage[latin9]{inputenc}
\usepackage{amstext}
\usepackage{graphicx}
\usepackage{amssymb}
\usepackage{esint}

\begin{document}
\title{Two-photon absorption and emission by Rydberg atoms in coupled cavities}

\author{Huaizhi Wu}
\email{huaizhi.wu@fzu.edu.cn} 
\author{Zhen-Biao Yang}
\email{zbyang@fzu.edu.cn} 
\author{Shi-Biao Zheng }
\affiliation{Department of Physics, Fuzhou University, Fuzhou 350002, People's Republic of China}


\begin{abstract}
We study the dynamics of a system composed of two coupled cavities, each interacting with a single Rydberg atom. The interplay between Rydberg-Rydberg interaction and photon hopping enables the transition of the atoms from the collective ground state to the double Rydberg excitation state by individually interacting with the optical normal modes and suppressing the up conversion process between them. The atomic transition is accompanied by the two-photon absorption and emission of the normal modes. Since the energy level structure of the atom-cavity system is photon number dependent there is only a pair of states being in the two-photon resonance. Therefore, the system can act as a quantum nonlinear absorption filter through the nonclassical quantum process, converting coherent light field into a non-classical state. Meanwhile, the vacuum field in the cavity inspires the Rydberg atoms to simultaneously emit two photons into the normal mode, resulting in obvious emission enhancement of the mode.
\end{abstract}

\maketitle
\section{Introduction}

The generation of nonclassical states of light has been a central
topic in quantum optics since the first demonstration of squeezed
states of light \cite{kimble_squeezing}. Quantum field in nonclassical
states reveals their nonclassical properties by exhibiting photon
anti-bunching, sub-Poisson photon-number statistics, and clearly negative
values of the Wigner function \cite{Haroche_book}. These states can
be used for understanding of quantum fluctuations beyond the standard
quantum noise limit and are essential sources in optical science and
engineering \cite{Yamamoto_MQO}. In this context, the two-photon
process, namely, the atoms transit from one energy level to another
through an intermediate energy level that simultaneously involves
two photons of the same frequency (the degenerate two-photon transition)
or of different frequencies (the nondegenerate two-photon transition),
has attracted great interest because it provides great opportunity
for producing light with nonclassical properties \cite{Bartzis_JOSAB91}. Indeed,
the two-photon absorption and emission are inherently nonclassical effects, which are expected to have potential applications in the realm of quantum techniques \cite{Hong_PRL86, Kwiat_PRL95, Simon_PRL05}.

Recently, high-finesse optical cavity has been used to couple Rydberg
atoms with quantized cavity modes, which presents potential applications
in studying photon nonlinearity and many-body physics \cite{Guerlin_PRA10_Cqed,Zhang_PRL13,Huang_PRA13}.
Neutral atoms excited by laser beams and the cavity field to high-lying
Rydberg states can interact through strong and long-range dipole-dipole
or van der Waals interaction \cite{Saffman_RMP10}. The quantum anharmonicity
of the energy level structure of the atom-cavity system enables the
study of two-photon absorption and three-photon absorption from a
probe beam \cite{Guerlin_PRA10_Cqed}. The optical nonlinearity has
been experimentally explored with strongly interacting Rydberg atoms
in cavities \cite{Parigi_PRL12}, even at the level of individual
quanta \cite{Peyronel_nature12}. Moreover, Zhang et al. have shown
that coupling of optical cavity to a lattice of Rydberg atoms can
be described by the Dicke model, the competition between the atom-atom
interaction and atom-light coupling can induce a novel superradiant
solid phase \cite{Zhang_PRL13}. On the other hand, rich quantum dynamics
has been found in the coupling of the coupled cavities with neutral
atoms \cite{Kim_PRA08_CC,Yang_epjd11}. Its potential applications
include realization of paradigmatic many-body models, such as the
Bose-Hubbard and the anisotropic Heisenberg models \cite{Hartmann_NP06}. 

Combine coupled cavities with interacting Rydberg atoms, a physical
model in analogous to the quantum dot-cavity coupling system, where
the cavity mode can be tuned to resonantly drive the two-photon transition
between the ground and the biexciton states, while the exciton states
are far-off resonance due to the biexciton binding energy \cite{Chen_PRL02_DotCavity},
will be discussed here. In the paper, we study the two-photon absorption
and emission process with two Rydberg atoms separately trapped in
coupled cavities. There exists two newly optical normal modes due to the
photon hopping between the two cavities. The collectively excited
energy level of the Rydberg atoms is shifted up or down according
to the sign of the F\"{o}rster defects, which induces the two-photon resonant
atomic transitions for either normal modes. In result, the blockade
of simultaneous excitation to the Rydberg state fails due to the photon
dynamics. The resonant transition frequency between the collective
ground state and the double Rydberg excitation state is photon number
dependent, leading to varied two-photon absorption rate for different
states of the cavity modes. The system can be used for realization
of quantum nonclassical processes and preparation of two-photon states.
The results are discussed in the context of micro-cavities, however,
the phenomenon may be found in other hybrid systems, such as Rydberg
atoms interacting with superconducting microwave devices \cite{Petrosyan_PRL08_MW,Hogan_PRL12_MW}.

\begin{figure}

\includegraphics[width=1\columnwidth]{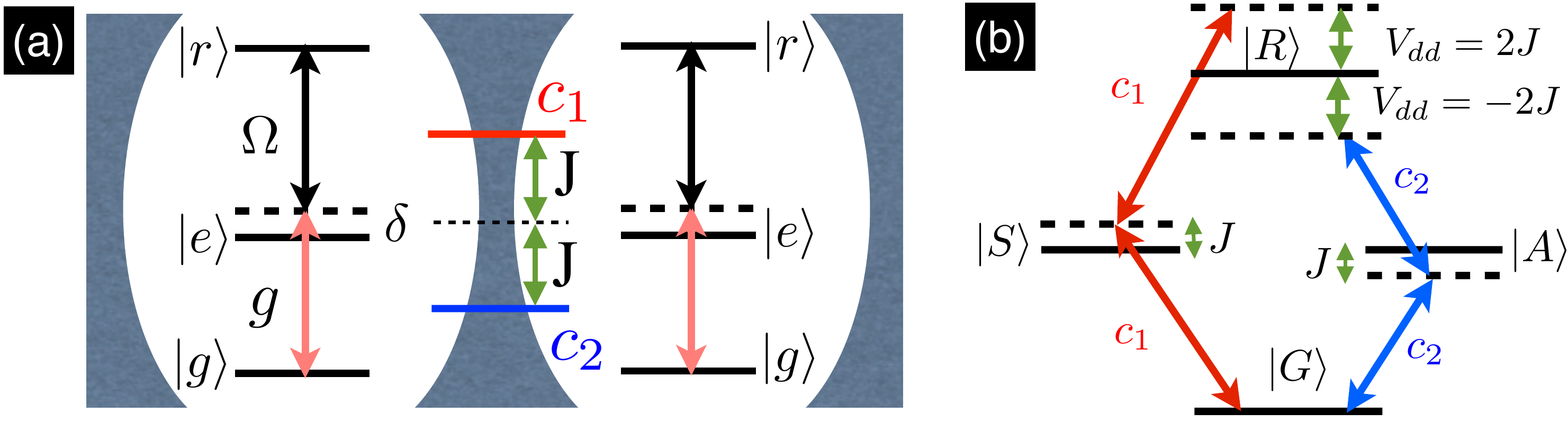}

\caption{(Color online) (a) Schematic setup: Two coupled cavities with each coupling to a single
Rydberg atom. Coherent Rydberg excitation between $|g\rangle$ and
$|r\rangle$ through a two-photon process via intermediate state $|e\rangle$.
The quantized cavity fields are coupled to the blue of the $|g\rangle\leftrightarrow|e\rangle$
transition, and laser light is tuned to the the red of the $|e\rangle\leftrightarrow|r\rangle$
transition. Photon hopping between the left and right cavity leads
to two new delocalized normal modes $c_{1}$ and $c_{2}$, with the bare
frequency separated by $2J$. (b) Effective model for heralded two-photon
transition between atomic collective states $|G\rangle$ and $|R\rangle$
through the intermediate symmetric and anti-symmetric entangled states
$|S\rangle$ and $|A\rangle$. The $c_{1}$ mode is in two-photon
resonance with the $|G\rangle\leftrightarrow|R\rangle$ transition
for $V_{dd}=2J$, while the $c_{2}$ mode is red detuned by $4J$.
For $V_{dd}=-2J$, the situation is in reverse.}
\label{fig:setup}
\end{figure}

\section{Interacting atoms in coupled cavities}
Consider the system composed of two coupled cavities, each interacting with
a Rubidium atom. This may be realized with micro-cavities (e.g. microtoroidal resonators), which
can couple to each other via the overlap of their evanescent  fields. The atoms have three relevant energy levels. As shown
in Fig. \ref{fig:setup}, the transition from $5S_{1/2}$ atomic ground state denoted
by $|g\rangle$ couples to a Rydberg excited level $|r\rangle$ through
a two-photon process via the $5P_{3/2}$ intermediate state $|e\rangle$
\cite{Guerlin_PRA10_Cqed}. The bare energies for the corresponding
energy levels are $\hbar\omega_{g}$, $\hbar\omega_{e}$ and $\hbar\omega_{r}$,
respectively. The atomic transitions $|g\rangle\leftrightarrow|e\rangle$
and $|e\rangle\leftrightarrow|r\rangle$ are coupled to quantized
cavity field of frequency $\omega_c$ and laser field of frequency $\omega$
with Rabi frequency $g$ and $\Omega$, respectively. The cavity field
is detuned by $\delta$ to the blue of the $|g\rangle\leftrightarrow|e\rangle$
transition, and laser beam is detuned by $\delta$ to the red of the
$|e\rangle\leftrightarrow|r\rangle$ transition. The energy shift
$V_{dd}$ for the collective atomic state $|r\rangle_{1}$$|r\rangle_{2}$
stemming from the Rydberg-Rydberg interaction prevents the simultaneous
excitation of the atoms to the Rydberg state $|r\rangle$. Photons
can hop between the left and right cavities with the rate $J$, giving
rise to a couple of optical normal modes with the frequencies $\omega_c\pm J$.
The Hamiltonian for the coupled atom-cavity system in the rotating
wave approximation (RWA) reads (assuming $\hbar=1$)
\begin{equation}
H=H_{c}+H_{a}+H_{af},
\end{equation}
with $$H_{c}=\omega_c(a_{1}^{\dagger}a_{1}+a_{2}^{\dagger}a_{2})+J(a_{1}^{\dagger}a_{2}+a_{1}a_{2}^{\dagger}), $$
\begin{eqnarray*}
H_{a}&=&\sum_{k=1,2}(\omega_{g}|g\rangle_{kk}\langle g|+\omega_{e}|e\rangle_{kk}\langle e|+\omega_{r}|r\rangle_{kk}\langle r|)\\
&+&V_{dd}|r\rangle_{1}|r\rangle_{22}\langle r|_{1}\langle r|,\nonumber 
\end{eqnarray*}
and $$H_{af}=\sum_{k=1,2}(g|e\rangle_{kk}\langle g|a_{k}+\Omega e^{-i\omega t}|r\rangle_{kk}\langle e|)+h.c.,$$
where $a_{k}(k=1,2)$ are annihilation operators for cavity fields
1 and 2, respectively. We have assumed that the coupling strengths
of the two atoms interacting with the respective local cavity modes
and laser beams are real and identical for simplicity. In the large
detuning regime, i.e., $\delta\gg\Omega,g$, the intermediate state
$|e\rangle$ will not be populated and can be adiabatically eliminated.
Thus, we have an effective Hamiltonian, which, in the interaction
picture, is given by
\begin{eqnarray}
H_{I} & = & H_{c}'+H_{a}'+H_{af}',
\label{eq:H_I1}
\end{eqnarray}
with
$$
H_{c}'=J(a_{1}^{\dagger}a_{2}+a_{1}a_{2}^{\dagger}),\nonumber \\
$$
$$
H_{a}'=V_{dd}|r\rangle_{1}|r\rangle_{22}\langle r|_{1}\langle r|,\nonumber \\
$$
and
\begin{eqnarray*}
H_{af}'&=&(\lambda\sum_{k=1,2}|r\rangle_{kk}\langle g|a_{k}+h.c.)+\lambda'\sum_{k=1,2}a_{k}^{\dagger}a_{k}|g\rangle_{kk}\langle g|\\&+&\lambda''\sum_{k=1,2}|r\rangle_{kk}\langle r|,
\end{eqnarray*}
where $\lambda=\Omega g/\delta$, $\lambda'=g^{2}/\delta$, and $\lambda''=\Omega^{2}/\delta$. 

For taking account of the Rydberg-Rydberg interaction between the
atoms, it is convenient to rewrite the atom-cavity interaction in
terms of the two-atom collective states \{$|G\rangle=|g\rangle_{1}|g\rangle_{2}$,
$|R\rangle=|r\rangle_{1}|r\rangle$, $|S\rangle=(|g\rangle_{1}|r\rangle_{2}+|r\rangle_{1}|g\rangle_{2})/\sqrt{2},$
$|A\rangle=(|g\rangle_{1}|r\rangle_{2}-|r\rangle_{1}|g\rangle_{2})/\sqrt{2}\}$,
and of the normal modes $c_{1}$ and $c_{2}$, which are symmetric
and antisymmetric superposition of the localized cavity annihilation
operators,
$$
c_{1}=\frac{1}{\sqrt{2}}(a_{1}+a_{2}),
$$
\begin{equation}
c_{2}=\frac{1}{\sqrt{2}}(a_{1}-a_{2}).
\end{equation}
Then, we have
\begin{eqnarray}
H_{I} & = & H_{c}'+H_{a}'+H_{af}',\label{eq:H_I2}
\end{eqnarray}
with
$$
H_{c}'=J(c_{1}^{\dagger}c_{1}-c_{2}^{\dagger}c_{2}),
$$
$$
H_{a}'=V_{dd}|R\rangle\langle R|,
$$
and
\begin{eqnarray*}
H'_{af} & = & [\lambda c_{1}(|S\rangle\langle G|+|R\rangle\langle S|)+\lambda c_{2}(|A\rangle\langle G|-|R\rangle\langle A|)\\
 & + & \frac{\lambda'}{2}(c_{1}^{\dagger}c_{2}+c_{2}^{\dagger}c_{1})|S\rangle\langle A|+h.c.]\\
 & + & (\frac{\lambda'}{2}(c_{1}^{\dagger}c_{1}+c_{2}^{\dagger}c_{2})+\lambda'')(|S\rangle\langle S|+|A\rangle\langle A|)\\
 & + & \lambda'(c_{1}^{\dagger}c_{1}+c_{2}^{\dagger}c_{2})|G\rangle\langle G|+2\lambda''|R\rangle\langle R|.
\end{eqnarray*}
The Hamiltonian $H_{c}'$ represents the non-interacting delocalized
modes with the frequency separated by $2J$. For the atom-field interaction
Hamiltonian $H_{af}'$, the first four terms describe the atomic transitions
from the collective ground state $|G\rangle$ to the double Rydberg
excitation state $|R\rangle$ through two independent channels $|G\rangle\rightarrow|S\rangle\rightarrow|R\rangle$
and $|G\rangle\rightarrow|A\rangle\rightarrow|R\rangle$, by interacting
individually with the normal modes $c_{1}$ and $c_{2}$. The two
transition channels link to each other via the optical frequency up
conversion associated with the atomic transition $|S\rangle\leftrightarrow|A\rangle$,
which is described by the fifth and sixth terms. The other terms are
stark shifts for the related collective atomic states.

\begin{figure}
\includegraphics[width=1\columnwidth]{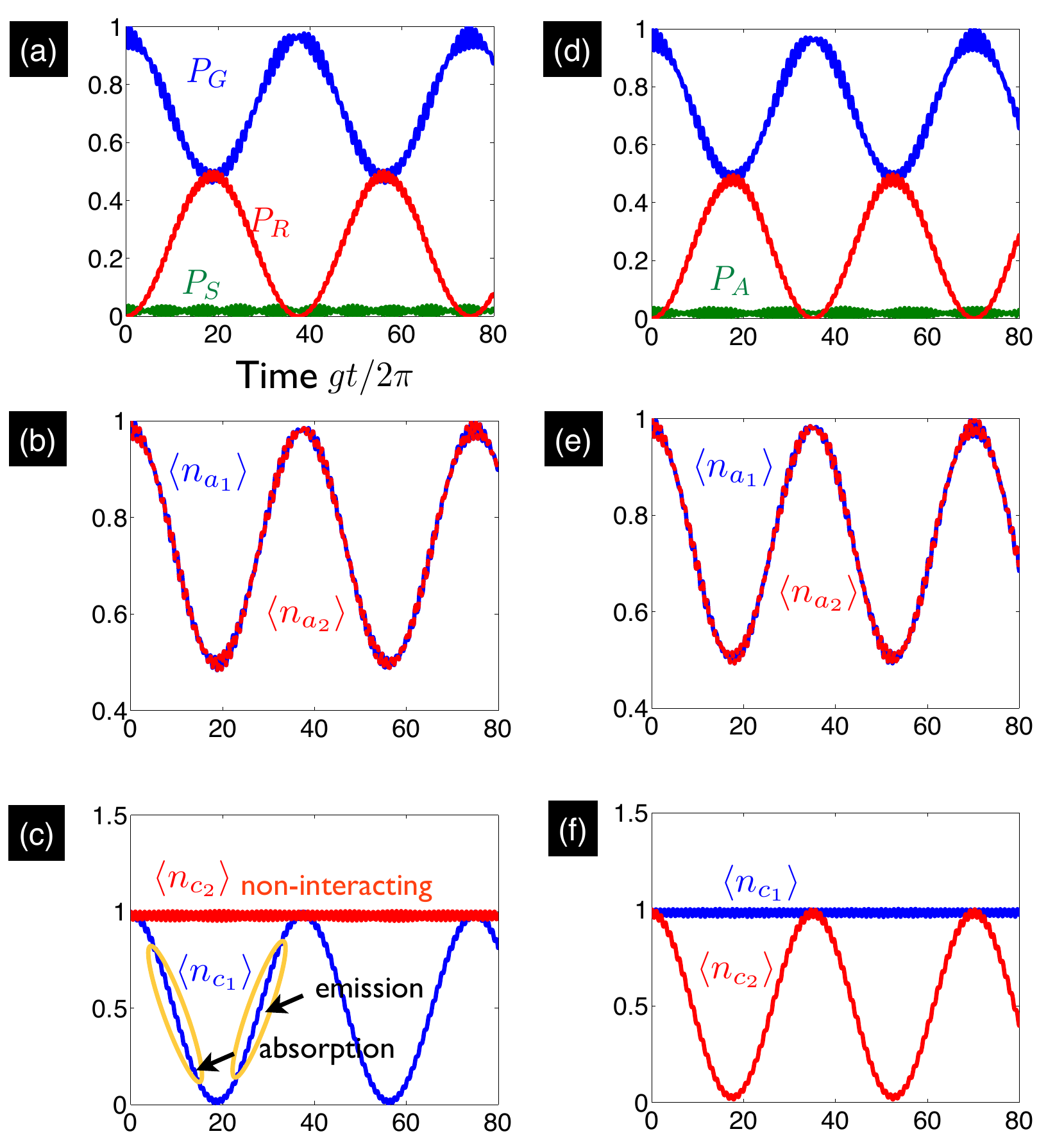}

\caption{\label{fig:dynamics}(Color online) Time-dependent population of the collective atomic
states, mean photon number for localized modes $a_{1}$ and $a_{2}$,
and mean photon number for delocalized normal modes $c_{1}$ and $c_{2}$
with the initial state $|G\rangle\otimes|1\rangle_{a1}|1\rangle_{a2}$.
The parameters are $\Omega/g=1$, $\delta=10g$, $J=10g$ with the
Rydberg-Rydberg interaction strength $V_{dd}=2J$ in (a)-(c) and $V_{dd}=-2J$
in (d)-(f). (a) and (d): Population of collective state $P_{G}$ and
$P_{R}$ display sinusoidal oscillation due to the absorption and
emission of two photons (see (b) and (e)). The photon dynamics for
$\langle a_{1}\rangle$ and $\langle a_{2}\rangle$ are identical,
while the dynamics of the delocalized modes $c_{1}$ and $c_{2}$
exhibit symmetric breaking. (c): For $V_{dd}=2J$, the absorption
of two photons from $c_{1}$ leads to atomic transition $|G\rangle\otimes|2\rangle_{c1}|0\rangle_{c2}\rightarrow|R\rangle\otimes|0\rangle_{c1}|0\rangle_{c2}$,
(f): while for $V_{dd}=-2J$, the atomic transition $|G\rangle\otimes|0\rangle_{c1}|2\rangle_{c2}\rightarrow|R\rangle\otimes|0\rangle_{c1}|0\rangle_{c2}$
is caused by absorption of two photons from $c_{2}$ .}
\end{figure}

\section{Two-photon coherent dynamics and Rydberg biexcitation}
To gain the insight of the full dynamics, we finally pass to a new
interaction Hamiltonian in a rotating frame with respect to $H_{c}'+H_{a}'=J(c_{1}^{\dagger}c_{1}-c_{2}^{\dagger}c_{2})+V_{dd}|R\rangle\langle R|$,
\begin{equation}
H_{I}'=H_{tr}+H_{st},
\end{equation}
with 
\begin{eqnarray*}
H_{tr} & = & \lambda c_{1}e^{-iJt}|S\rangle\langle G|+\lambda c_{2}e^{iJt}|A\rangle\langle G|\\
 & + & \lambda c_{1}e^{i(V_{dd}-J)t}|R\rangle\langle S|-\lambda c_{2}e^{i(V_{dd}+J)t}|R\rangle\langle A|\\
 & + & \frac{\lambda'}{2}(c_{1}^{\dagger}c_{2}e^{i2Jt}+c_{2}^{\dagger}c_{1}e^{-i2Jt})|S\rangle\langle A|+h.c.,
\end{eqnarray*}
and
\begin{eqnarray*}
H_{st}&=&(\frac{\lambda'}{2}(c_{1}^{\dagger}c_{1}+c_{2}^{\dagger}c_{2})+\lambda'')(|S\rangle\langle S|+|A\rangle\langle A|) \\
&+&\lambda'(c_{1}^{\dagger}c_{1}+c_{2}^{\dagger}c_{2})|G\rangle\langle G|+2\lambda''|R\rangle\langle R|.
\end{eqnarray*}
(i) Without considering the photon number dependent stark shifts $H_{st}$.
The normal mode $c_{1}$ is blue-detuned by $J$ from the transition
$|G\rangle\leftrightarrow|S\rangle$ and $c_{1}$ is red-detuned by
$J$ from the transition $|G\rangle\leftrightarrow|A\rangle$, see
Fig. 1(b). To decouple the transition channels $|G\rangle\rightarrow|S\rangle\rightarrow|R\rangle$
and $|G\rangle\rightarrow|A\rangle\rightarrow|R\rangle$, the optical
frequency up conversion should be suppressed. This can be met if the
frequency separation of the normal modes is much greater than the
conversion rate (dispersive regime), i.e., 2$J\gg\sqrt{n_{c1}n_{c2}}\lambda'$/2.
Note that the sign of the Rydberg-Rydberg interaction strength is
determined by the sign of the energy gap in the F\"{o}rster process \cite{Walker_PRA08_Zeeman}.
Now if $V_{dd}=2J$, the atomic transition $|G\rangle\leftrightarrow|R\rangle$
mediated by symmetric entangled state $|S\rangle$ is in resonance
with twice the photon frequency of the normal mode $c_{1}$, while
the other channel $|G\rangle\rightarrow|A\rangle\rightarrow|R\rangle$
is out of resonance and is detuned by $4J$. Therefore, under the
condition $J\gg\sqrt{n_{c1}}\lambda$, we can finally obtain an effective
Hamiltonian by using the time averaging approach to describe this
two-photon transition process \cite{James_effectiveH}, 
\begin{equation}
H_{eff}=\xi(|G\rangle\langle R|c_{1}^{\dagger2}+|R\rangle\langle G|c_{1}^{2}),\label{eq:H_eff1}
\end{equation}
where $\xi=\lambda^{2}/J$, and the stark shift terms $(\lambda^{2}/J)[(c_{1}^{\dagger}c_{1}-c_{2}^{\dagger}c_{2})|G\rangle\langle G|+|R\rangle\langle R|(c_{1}c_{1}^{\dagger}+c_{2}c_{2}^{\dagger}/3)]$
have been neglected because they are much less than the photon number
dependent energy $H_{st}$ . While if $V_{dd}=-2J$, the transition
channel $|G\rangle\leftrightarrow|R\rangle$ mediated by the singlet
state $|A\rangle$ is in resonant with twice the frequency of the
normal mode $c_{2}$, and the channel $|G\rangle\rightarrow|S\rangle\rightarrow|R\rangle$
related to $c_{1}$ is out of resonance. The effective Hamiltonian
is then otherwise given by 
\begin{equation}
H_{eff}=\xi(|G\rangle\langle R|c_{2}^{\dagger2}+|R\rangle\langle G|c_{2}^{2}).\label{eq:H_eff2}
\end{equation}
Thus, the blockade of the double Rydberg excitations may be wrecked
due to the photon hopping through two-photon absorption. (ii) Two-photon
transition including $H_{st}$. Taking stark shift $H_{st}$ into
consider, the two-photon resonant transition may break down if $|G\rangle$
and $|R\rangle$ are shifted by different amount depending on the
photon number of the normal modes. Set $\Omega=g$, this implies the two-photon
resonance condition $V_{dd}=2J+(\langle n_{c1}\rangle-2)\lambda$
and $V_{dd}=-2J+(\langle n_{c2}\rangle-2)\lambda$ for $c_{1}$ mode
and $c_{2}$ mode, respectively, when the atoms are initially in the
ground state $|g\rangle$. The energy shifts of the symmetric and
anti-symmetric entangled atomic states will slightly modify the effective
two-photon coupling rate.

\begin{figure}
\includegraphics[width=1\columnwidth]{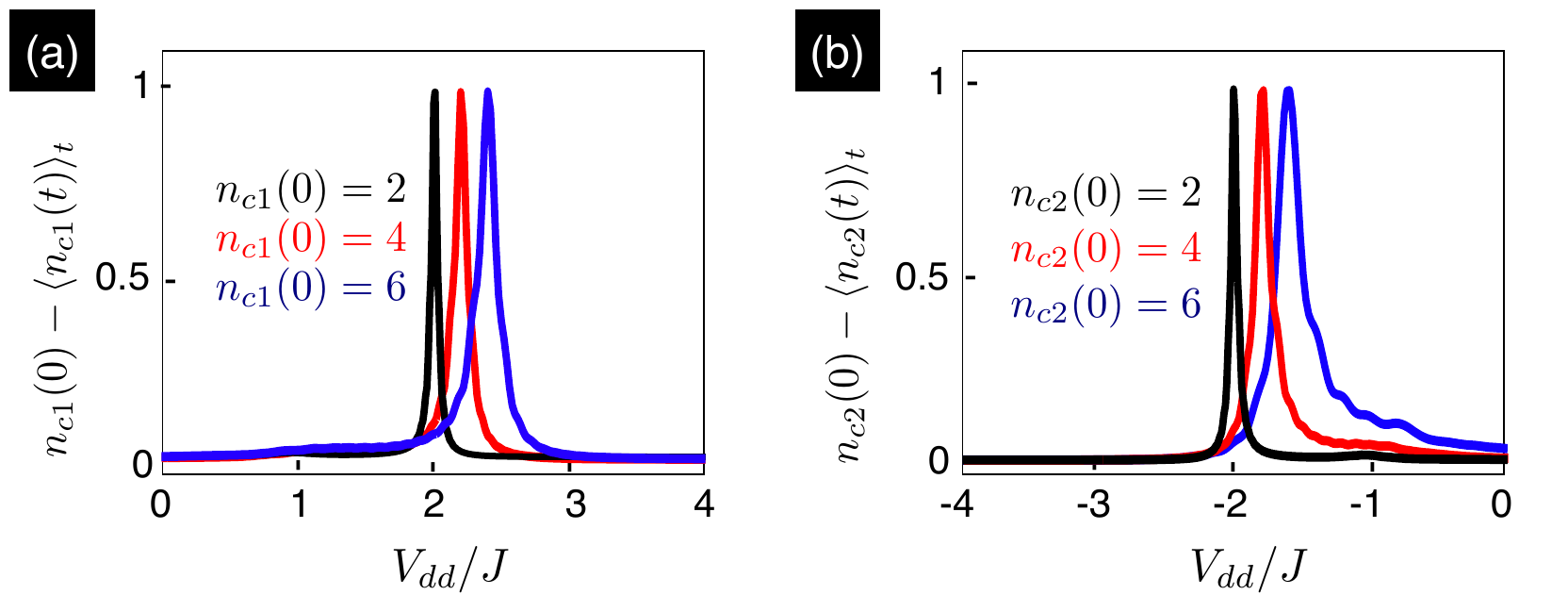}

\caption{(Color online) (a) Time-averaged photon absorption versus $V_{dd}/J$
with initial system state (a) $|G\rangle\otimes|n_{c1}\rangle_{c1}|0\rangle_{c2}$
and (b) $|G\rangle\otimes|0\rangle_{c1}|n_{c2}\rangle_{c2}$ over
$t\in[0,2\pi/\sqrt{n_{c1,2}(n_{c1,2}-1)}\xi]$. The absorption centers
are shifted according to the photon number dependent Stark shifts
described by $H_{st}$. Other parameters as in Fig. 2.}
\label{fig:MPhot_vdd}
\end{figure}

The coherent quantum dynamics in this atom-cavity coupled system can
be read from Fig.\ref{fig:dynamics}, in which we have shown the time-dependent
population of the collective atomic states with the system initial
state $|G\rangle\otimes|1\rangle_{a1}|1\rangle_{a2}$. In terms of
the delocalized normal modes, the initial state of the cavity fields
can be rewritten as $a_{1}^{\dagger}a_{2}^{\dagger}|0\rangle_{a1}|0\rangle_{a2}=\frac{1}{2}(c_{1}^{\dagger}+c_{2}^{\dagger})(c_{1}^{\dagger}-c_{2}^{\dagger})|0\rangle_{c1}|0\rangle_{c2}=(|2\rangle_{c1}|0\rangle_{c2}-|0\rangle_{c1}|2\rangle_{c2})/\sqrt{2}$.
In this case, the energy shifts for $|G\rangle$ and $|R\rangle$
are both $2\lambda$, which guarantee the two-photon resonance condition.
In Fig. \ref{fig:dynamics}(a) and \ref{fig:dynamics}(d), the Rabi
oscillation between $|G\rangle$ and $|R\rangle$ clearly demonstrates
the photon absorption and emission processes, and the excitation of
the symmetric and anti-symmetric entangled states are well suppressed.
The probability for detecting $|R\rangle$ can only reach 0.5 or so
in each plot because the transition channels via the intermediate
state $|S\rangle$ and $|A\rangle$ are selected by the sign of the
energy shift $V_{dd}$. In this process, the photon dynamics of the
localized modes $a_{1}$ and $a_{2}$ display exactly the same behavior
and remain symmetric. It means that the two localized photons are
absorbed and emitted simultaneously all the time (see Fig. \ref{fig:dynamics}(b)
and \ref{fig:dynamics}(e)). In contrast, the coupling to the delocalized normal
modes are symmetry breaking. Either the transition between $|G\rangle\otimes|2\rangle_{c1}|0\rangle_{c2}$
and $|R\rangle\otimes|0\rangle_{c1}|0\rangle_{c2}$ with $V_{dd}=2J$
or that between $|G\rangle\otimes|0\rangle_{c1}|2\rangle_{c2}$ and $|R\rangle\otimes|0\rangle_{c1}|0\rangle_{c2}$
with $V_{dd}=-2J$ happens. That is accompanied by the absorption
and emission of two delocalized photons (Fig.\ref{fig:dynamics}(c)
and \ref{fig:dynamics}(f)). Note that the effective coupling strength
should be revised by $\xi'=\lambda^{2}/(J\pm\lambda/2)$ for $V_{dd}=\pm2J$
due to the energy shifts of $|S\rangle$ and $|A\rangle$, which leads
to the slightly different time period of oscillation for $\langle n_{c1}\rangle$
and $\langle n_{c2}\rangle$. The time evolution of the system dynamics
discussed above can be analytically calculated by solving the Schr\"{o}dinger
equation $i\hbar\dot{\psi}(t)=H_{eff}\psi(t)$. Without loss of generality,
we focus on the two-photon transition with respect to the delocalized
mode $c_{1}$ described by Eq. (\ref{eq:H_eff1}), from which we can
obtain the quantum state of the system at time $t$, 
\begin{eqnarray}
|\psi(t)\rangle&=&\frac{1}{\sqrt{2}}(C_{g}(t)|G\rangle\otimes|2\rangle_{c1}|0\rangle_{c2}+C_{r}(t)|R\rangle\otimes|0\rangle_{c1}| 0\rangle_{c2}) \nonumber \\
&-&\frac{1}{\sqrt{2}}|G\rangle\otimes|0\rangle_{c1}|2\rangle_{c2}.
\end{eqnarray}
 In addition, by appropriately choosing the interaction time, the
system will evolve from $|G\rangle\otimes(|2\rangle_{c1}|0\rangle_{c2}-|0\rangle_{c1}|2\rangle_{c2})/\sqrt{2}$
($|G\rangle\otimes|1\rangle_{a1}|1\rangle_{a2}$) onto $|G\rangle\otimes(|2\rangle_{c1}|0\rangle_{c2}+|0\rangle_{c1}|2\rangle_{c2})/\sqrt{2}$
($|G\rangle\otimes(|2\rangle_{a1}|0\rangle_{a2}+|0\rangle_{a1}|2\rangle_{a2})$, which is a two-photon NOON state for localized modes.

\begin{figure}
\includegraphics[width=1\columnwidth]{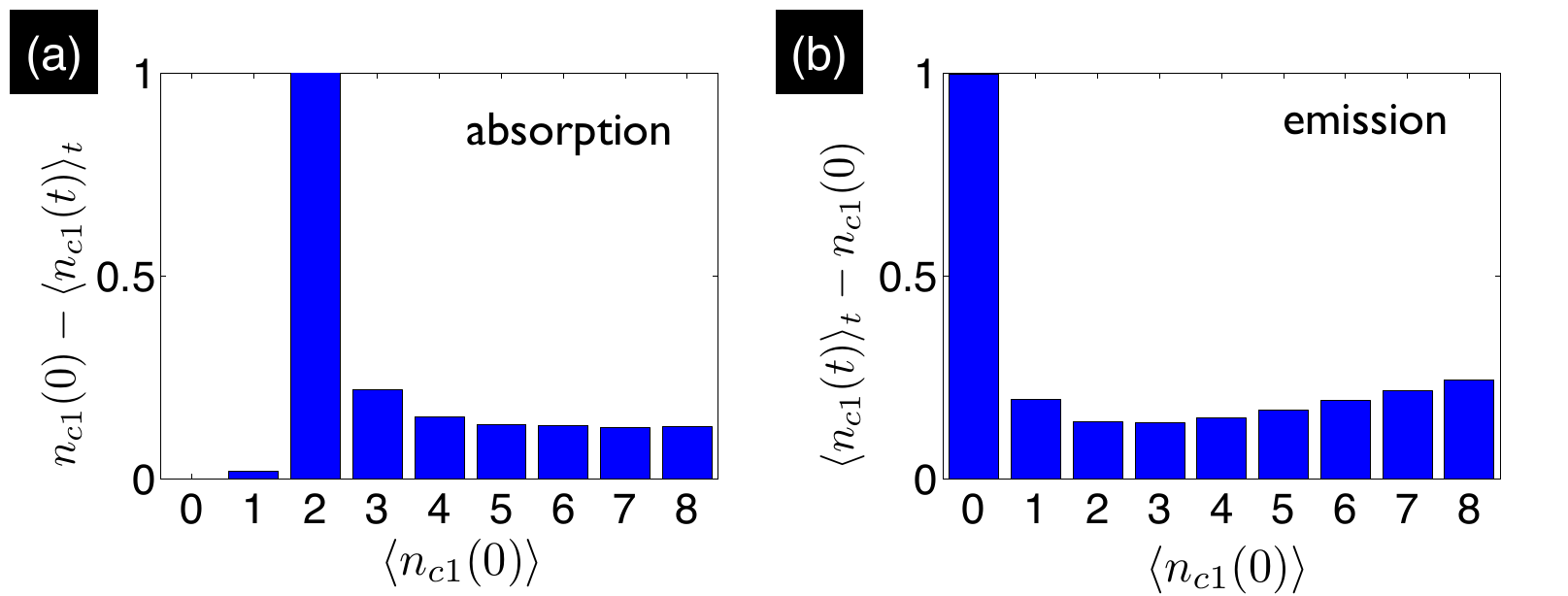}

\caption{(Color online) (a) Time-averaged photon absorption in $t\in[0,2\pi/\sqrt{2}\xi]$
versus $n_{c1}$ with initial system state $|G\rangle\otimes|n_{c1}\rangle_{c1}|0\rangle_{c2}$.
(b) Time-averaged photon emission in $t\in[0,2\pi/\sqrt{2}\xi]$ versus
$n_{c1}$ with initial system state $|R\rangle\otimes|n_{c1}\rangle_{c1}|0\rangle_{c2}$.
Other parameters as in Fig. 3.}
\label{fig:mean_photon}
\end{figure}

The physical model can be further understood by looking into the time
averaged photon absorption $n_{c1}(0)-\langle n_{c1}(t)\rangle_{t}$
($n_{c2}(0)-\langle n_{c2}(t)\rangle_{t}$) as a function of the Rydberg-Rydberg
interaction strength for the initial state $|G\rangle\otimes|n_{c1}\rangle_{c1}|0\rangle_{c2}$
($|G\rangle\otimes|0\rangle_{c1}|n_{c2}\rangle_{c2}$) with varied
photon number (see Fig. \ref{fig:MPhot_vdd}). It is found that the
two-photon absorption centers are shifted according to the photon
number dependent energy shifts for $|G\rangle\otimes|n_{c1}\rangle_{c1}|0\rangle_{c2}$
and $|G\rangle\otimes|0\rangle_{c1}|n_{c2}\rangle_{c2}$ that are
illustrated by $H_{st}$. For $V_{dd}\simeq2J$ or $V_{dd}\simeq-2J$,
only when $n_{c1}=2$ and $n_{c2}=0$, or $n_{c1}=0$ and $n_{c2}=2$,
the interplay between Rydberg-Rydberg interaction and photon tunneling
will cause absorption of two photons from the normal modes. The resonant
two-photon transition gives time-averaged photon absorption of one.
For the initial state being $|G\rangle\otimes|n_{c1}\rangle_{c1}|0\rangle_{c2}$
and $n_{c1}\neq2$, this corresponds to the dispersive interaction
regime because the $|G\rangle\leftrightarrow|R\rangle$ transition
is detuned by $(n_{c1}-2)\lambda$, which is much larger than the
effective $|G\rangle\leftrightarrow|R\rangle$ coupling strength $\sqrt{n_{c1}(n_{c1}-1)}\xi$.
As shown in Fig. \ref{fig:mean_photon}(a), the photon absorption becomes
weaker and weaker as the initial photon number of $c_{1}$ mode increases.
It is also interesting to study the time averaged photon emission
$\langle n_{c1}(t)\rangle_{t}-n_{c1}(0)$ with the system initial
state $|R\rangle\otimes|n_{c1}\rangle$ (see Fig. \ref{fig:mean_photon}(b)).
Here, the atoms are both initially in the Rydberg excited state. The
atomic transition $ $$|R\rangle\rightarrow|G\rangle$ regularly happens
accompanied by simultaneously emitting two photons for the normal
mode being in the vacuum state, giving rise to emission enhancement
of the mode. For $n_{c1}(t=0)>1$, the normalized photon emission
given by $(\langle n_{c1}(t)\rangle_{t}-n_{c1}(0))/n_{c1}(0)$ goes
down as $n_{c}(0)$ grows.

\begin{figure}
\includegraphics[width=1\columnwidth]{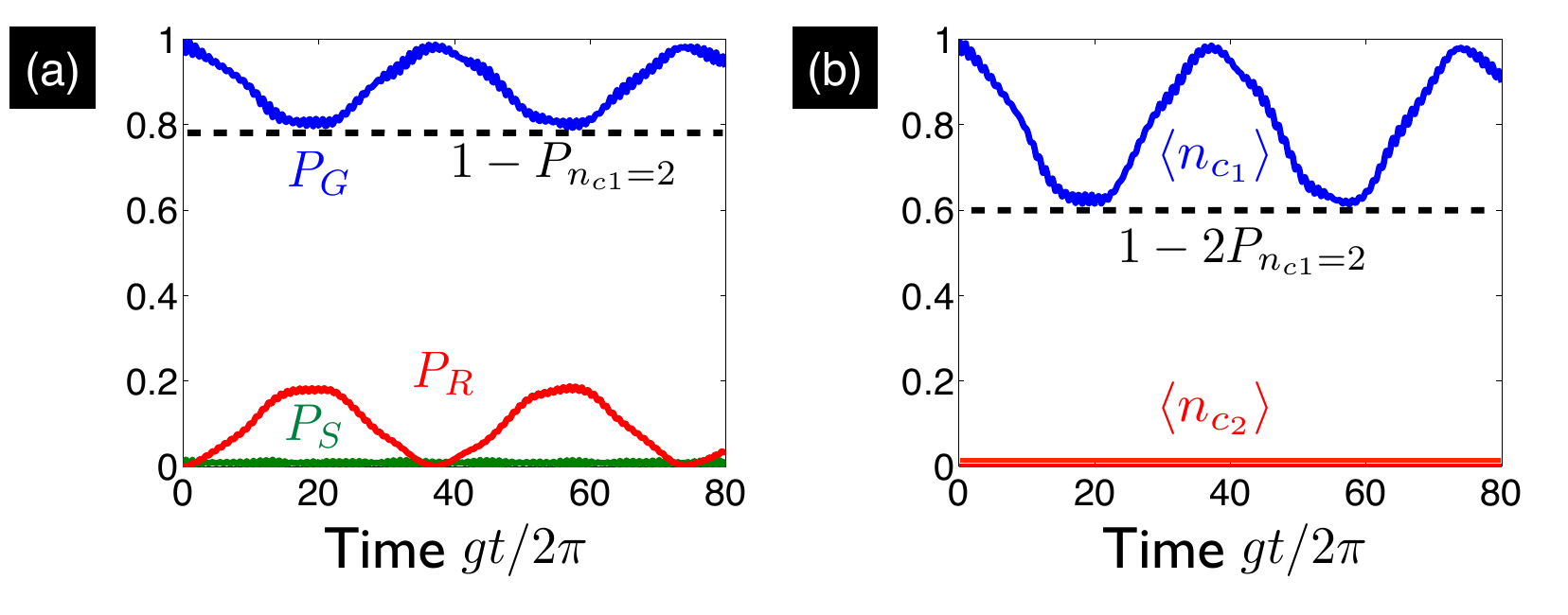}

\caption{(Color online) Time dependent population of the collective atomic
states and mean photon number for optical normal modes $c_{1}$ and
$c_{2}$ for initially the atoms in the state $|G\rangle$ and the
cavity modes in the coherent states $|\alpha\rangle_{a1}|\beta\rangle_{a2}=|(\alpha+\beta)/\sqrt{2}\rangle_{c1}|(\alpha-\beta)/\sqrt{2}\rangle_{c2}$,
where $\alpha=\beta=1/\sqrt{2}$. The interaction of the atoms with
cavity fields is mainly dominated by the $|G\rangle\otimes|2\rangle_{c1}|0\rangle_{c2}\rightarrow|R\rangle\otimes|0\rangle_{c1}|0\rangle_{c2}$
transition. The oscillational amplitude is limited by the probability
amplitude of the component $|2\rangle_{c1}$ for $|(\alpha+\beta)/\sqrt{2}\rangle_{c1}$
expanded in Fock space. The normal mode $c_{2}$ is unpopulated during
the interaction. The parameters are $\Omega=g$, $\delta=10g$, $J=0.998g,$
and $V_{dd}=2g$. }
\label{fig:coherent}
\end{figure}

\section{Quantum nonclassical  process and Quantum filter}
Now we consider the localized cavity fields that are initially in
the coherent states $|\alpha\rangle_{a1}$ and $|\beta\rangle_{a2}$
respectively. The quantum state of the localized two-mode field can
be rewritten as $|\alpha\rangle_{a1}|\beta\rangle_{a2}=|(\alpha+\beta)/\sqrt{2}\rangle_{c1}|(\alpha-\beta)/\sqrt{2}\rangle_{c2}$
in terms of the normal modes $c_{1}$ and $c_{2}$, which are the
coherent states of mean photon number $\langle N_{c1}\rangle=|\alpha+\beta|^{2}/2$
and $\langle N_{c2}\rangle=|\alpha-\beta|^{2}/2$. For $\alpha=\beta$,
the $c_{2}$ mode is in the vacuum state. Thus, if the atoms are both
in the ground state $|g\rangle$ and the Rydberg-Rydberg interaction
strength is $V_{dd}=2J$, only the transition channel $|G\rangle\rightarrow|S\rangle\rightarrow|R\rangle$
will be opened for the coupling of $|G\rangle$ with $|R\rangle$.
We can then simply focus on the photon dynamics of the $c_{1}$ mode.
To study the $c_{2}$ mode, we can alternatively set $\alpha=-\beta$.
Without loss of generality, we will assume $\alpha=\beta$ in the
following. The $c_{1}$ mode can be expanded in the Fock state representation
as $|(\alpha+\beta)/\sqrt{2}\rangle_{c1}=exp(-|\alpha+\beta|^{2}/2)\sum_{n=0}^{\infty}(((\alpha+\beta)/\sqrt{2})^{n}/\sqrt{n!})|n\rangle_{c1}$.
As discussed above, the effective Hamiltonian in Eq.(\ref{eq:H_eff1})
or Eq.(\ref{eq:H_eff2}) holds only when there are two photons in
$c_{1}$ mode. Otherwise, the collective atomic states $|G\rangle$
and $|R\rangle$ will be shifted by different amount and the $|G\rangle\leftrightarrow|R\rangle$
transition is thus out of resonance. This can be used for demonstration
of the quantum nonclassical process \cite{Saleh_PRL13_QPNonclassical},
where a coherent state will be transformed to a nonclassical state.
The coherent dynamics of the Rydberg atoms interacting with coherent
cavity fields is shown in Fig. \ref{fig:coherent}. The oscillation
is mainly due to the interplay of $|G\rangle\otimes|0\rangle_{c1}|2\rangle_{c2}$
with $|R\rangle\otimes|0\rangle_{c1}|0\rangle_{c2}$ weakly perturbed
by $|G\rangle\otimes|n\rangle_{c1}|0\rangle_{c2}$ ($n_{c1}\neq2$).
Therefore, the minimum of the atomic population $P_{G}$ and the mean
photon number of $c_{1}$ mode $\langle n_{c1}\rangle$ are approximately
given by ($1-P_{n_{c1}=2}$) and ($1-2P_{n_{c1}=2}$), respectively,
with $P_{n_{c1}=2}$ the probability amplitude of $|2\rangle_{c1}$
in the expansion of coherent state in the Fock space.

\begin{figure}
\includegraphics[width=1\columnwidth]{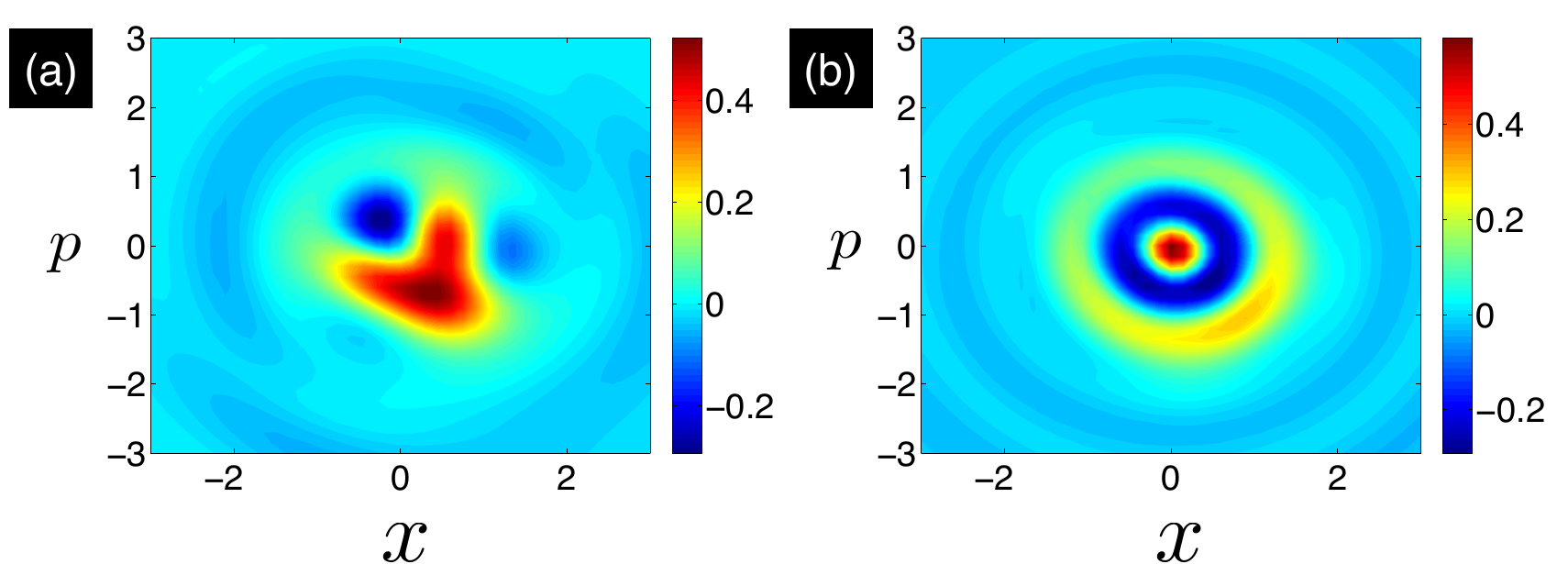}

\caption{(Color online) Wigner function $W(x,p)$ of the normal mode $c_{1}$ after
interaction with Rydberg atoms. The initial states are: (a) $|G\rangle\otimes|\alpha=1/\sqrt{2}\rangle_{a1}|\beta=1/\sqrt{2}\rangle_{a2}$
and (b) $|R\rangle\otimes|\alpha=1/\sqrt{2}\rangle_{a1}|\beta=1/\sqrt{2}\rangle_{a2}$.
Other parameters as in Fig. 5.}
\label{fig:Wigner}
\end{figure}

Using this system, we can realize a quantum nonlinear absorption filter
and the nonclassical quantum optical process defined by Rahimi-Keshari
et al. \cite{Saleh_PRL13_QPNonclassical}. For the atoms interacting
with coherent cavity fields with the initial state $|G\rangle\otimes|\sqrt{2}\alpha\rangle_{c1}|0\rangle_{c2}$,
the measurement of the atoms in the state $|G\rangle$ at $t=\pi/2\sqrt{2}\xi$
will collapse the normal mode $c_{1}$ into a nonclassical quantum
state finally by removing the component Fock state $|2\rangle$
from coherent $c_{1}$ mode. 
Note that the density operator of an arbitrary field state can be written as $\rho=\int P(\alpha)|\alpha\rangle\langle\alpha|d^{2}\alpha$ by means of a diagonal representation in terms of the coherent states, and the probability distribution of the photon-number states in $\rho$
is given by $p(n)=\int P(\alpha)|\langle n|\alpha\rangle|^{2}d^{2}\alpha$. Because $|\langle n|\alpha\rangle|^{2}>0$, $p(n)$ can not be zero for any $n$ when $P(\alpha)$ is a true probability density. Then, any field state for which $p(n)=0$ has no classical analog and deserves special attention \cite{Mandel_and_Wolf}. 
Therefore, the state of the normal mode $c_{1}$ with $p(n=2)=0$ is purely quantum mechanical.
The Wigner function for such kind of nonclassical states is shown in Fig.\ref{fig:Wigner}(a), from which
we clearly see the negative value close to the origin demonstrating
its nonclassical nature. The component Fock state $|2\rangle_{c1}$
has been absorbed by the atoms being excited to $|R\rangle$. 
It corresponds to a quantum filter for a special photon number state that has been realized with linear optics based on multi-photon interference and measurement induced amplitude nonlinearity \cite{Sanaka_PRL06,Bartley_PRA2012}.
On the other hand, while the atoms initially in the collective state $|R\rangle$
interact with the coherent fields, ideally, the measurement of the
atoms in the collective state $|G\rangle$ will collapse the normal
mode $c_{1}$ into the Fock state $|2\rangle_{c1}$, which is weakly
influenced by the components of the other Fock states. The Wigner
function of the normal mode $c_{1}$ after atomic measurement is shown
in Fig. \ref{fig:Wigner}(b), which predicates a Fock state $|2\rangle_{c1}$
in despite of tiny distortion. This can probably be used for realization
of two-photon source, in particular, for the cavity fields initially
being in the vacuum states. Although we assumed the cavity fields
are initially in the coherent states, the characteristics of current
model are applicable to tailor different quantum states of light,
such like thermal fields in coupled cavities.

\section{The experimental realization}
The schematic setup and the theoretical model studied in this paper may be experimentally realized with quantum optical devices such as coupled toroid microcavities \cite{Grudinin_prl10_coupledtoroid} or waveguide-coupled Fabry-Perot cavities \cite{Lepert_njp11_coupledcav,Lepert_arxiv2013}. We assumed that each cavity contains only a single optical mode. The single mode assumption is valid when the cavity coupling $J$
  is small compared with the free spectral range (FSR) of the each uncoupled cavity. With toroid microcavities, the coupling of the initially uncoupled whispering-gallery modes can be realized via control of the air gap between the microtoroids, and therefore the overlap of evanescent fields. This has been demonstrated recently by Grudinin et al. \cite{Grudinin_prl10_coupledtoroid}. The cavity coupling $J$
  can be tuned ranging from $5$MHz to $5$GHz, which is much less than the FSR that is on the order of several hundred GHz \cite{Spillane_CaltechThesis04}. On the other hand, the optical modes can be made degenerate in frequency by thermal control of the microtoroids \cite{Grudinin_prl10_coupledtoroid}. Therefore, only two modes (one from each microtoroid) contribute to the coupled system. The atoms coupled to the resonators' evanescent field can then interact through dipole-dipole or van der Waals interaction \cite{Aoki_Nature06}. For waveguide-coupled Fabry-Perot cavities, the cavity coupling of the strength $J\sim2\pi\times50$MHz and the free spectral range beyond $2\pi\times1$GHz are achievable \cite{Lepert_njp11_coupledcav}. Therefore, the single mode cavity assumption proposed here should be reasonable. In addition, the waveguide-coupled Fabry-Perot cavities setup was studied in detail for its potential application in constructing a Jaynes-Cumming lattice and simulating the quantum dynamics of a spin chain \cite{Lepert_njp11_coupledcav}. The microcavities are open in the transverse direction and the longitudinal cavity axes are separated by several microns, giving access to lasers that excite the atoms to the Rydberg state. Assuming that each atom is at the center of its cavity, the Rydberg coupling then depends on the transverse distance between the microcavities. Therefore, the Rydberg-Rydberg interaction should in principle be the same as the Rydberg atoms in vacuum.

For experimental demonstration of the two-photon absorption and emission,
the parameter $\xi$ of the effective Hamiltonian should be much larger
than effective decoherence rates via the photons and the excited states.
Set $\Omega=g$, $\delta=10g$, and $J=g$, then we have $\lambda=\Omega g/\delta=0.1g$,
$\xi=\lambda^{2}/J=0.01g$, and the time needed for preparing nonclassical
states shown in Fig. \ref{fig:Wigner} is $t=\pi/2\sqrt{2}\xi\simeq1.11\times10^{2}g^{-1}$.
Since the intermediate state $|e\rangle$ is off-resonantly coupled
with the ground state $|g\rangle$ and the Rydberg state $|r\rangle$
, the effective decay rate for $|e\rangle$ is $\gamma_{e}=(g^{2}/\delta^{2}+\Omega^{2}/\delta^{2})\gamma$
for $\delta\gg\Omega,g,\gamma$, where $\gamma$ is the spontaneous
emission rate. The Rydberg state with principal quantum number $n=70$
has a spontaneous decay rate $\gamma_{r}=2\pi$$\times0.55$kHz that
is much smaller than $\gamma$. The cavity decay rate should fulfill
the condition $\kappa\ll\xi\sim0.01J$, which implies the photons
fast tunnel to next cavity before decay into free space. These decoherence
sources will induce intrinsic errors for the implementation. On the
other hand, the Rydberg-Rydberg interaction arises from the intrinsic
F\"{o}rster interaction, which can lead to the energy shift $V_{dd}$
approximating to $200$MHz with the interatomic distance around $7$$\mu m$
\cite{Guerlin_PRA10_Cqed}. The sign of the energy shift is determined
by the sign of F\"{o}rster defects correlated with the selected transition
channels \cite{Walker_PRA08_Zeeman}. The parameter regime above can
be achieved with the micro-cavities, where atom-cavity interacting
system with the cooperativity factor as high as $C=g^{2}/2\kappa\gamma\sim10^{5}$
is predicted to be available \cite{Spillane_PRA05_HighQ}. The micro-cavities
with the size tens of $\mu m$ can be coupled via the overlap of their
evanescent fields. Set $\kappa\sim10^{-3}g$ and $\gamma\sim10^{-3}g$,
the errors of the prepared nonclassical state due to decoherence approximates
to $E\simeq(\gamma_{e}+\gamma_{r}+\kappa)t\simeq0.12$. Without seeing
a quantum jump from the coupled cavities, the dissipative dynamics
of the system can be described by the non-Hermitian Hamiltonian $H_{NH}=H_{I}-i\kappa/2(a_{1}^{\dagger}a_{1}+a_{2}^{\dagger}a_{2})$,
using which we find that the successful probability of the current
proposal decreases according to the exponential factor $e^{-\bar{n}_{c}\kappa t}$
with $\bar{n}_{c}$ the mean photon number in coupled cavities, while
the Wigner function for the prepared state is only slightly changed.
To improve the fidelity, an atomic ensemble acting as a two-level
``superatom'' can be placed inside the cavities, instead of a single
Rydberg atom \cite{Vuletic_NP2006_superatom}.

\section{Conclusion}
In conclusion, we have studied the interaction between Rydberg atoms
and the normal modes in coupled cavities. The dispersive atom-cavity
interaction results in nonlinear electronic-level shifts depending
on the photon-number of the normal modes. The Rydberg atoms can simultaneously
absorb (emit) two photons from (into) one of the normal modes relying
on the sign of the Rydberg-Rydberg interaction induced energy shift.
There is only one transition channel that is in two-photon resonance,
which can be used for generation of nonclassical states of light and realization of a quantum filter. The physical realization of this scheme can
be realized with coupled micro-cavities, however, the alternative
candidates of experimental setup include ultrahigh-Q coupled nano-cavity
based on photonic crystals \cite{Notomi_NPhoton08} and superconducting
microwave devices \cite{Petrosyan_PRL08_MW,Hogan_PRL12_MW}. This
scheme promises a new avenue for manipulation of quantum state of
light and realization of nonclassical quantum optical process.

\acknowledgments
This work was supported by the Major State Basic Research Development Program of China under grant no. 2012CB921601, the National Natural Science Foundation of China under grant no. 11247283, no. 11305037, and no. 11374054, the Natural Science Foundation of Fujian Province under grant no. 2013J01012, and the fund from Fuzhou University.
\bibliographystyle{apsrev}
\bibliography{/Users/huaizhiwu/Lab_Huaizhi/Research_Projects/Coupled_cavity_with_Rydberg_atoms/reference/Ryd_CoupC_citations}

\end{document}